\documentclass[letterpaper,preprint,preprintnumbers,amsmath,amssymb]{revtex4}
\usepackage{graphicx}% Include figure files
\usepackage{dcolumn}% Align table columns on decimal point
\usepackage{bm}% bold math
\usepackage[left]{lineno}
\usepackage[usenames]{color}

\begin{document}

\includegraphics*[viewport=90 50 600 730, page=1]{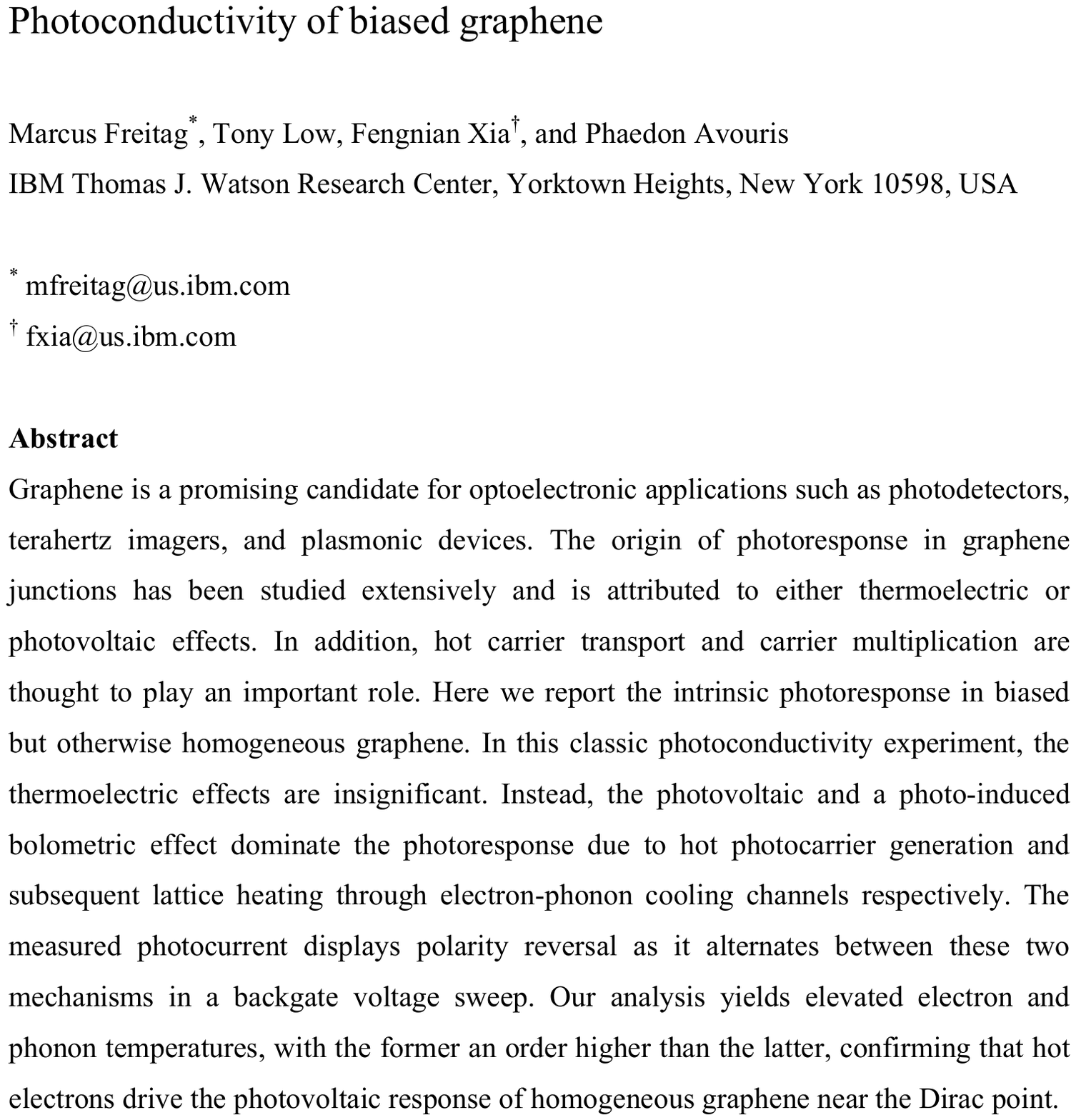}
\includegraphics*[viewport=80 50 600 730, page=2]{Gr_Photoconductivity.pdf}
\includegraphics*[viewport=80 50 600 730, page=3]{Gr_Photoconductivity.pdf}
\includegraphics*[viewport=80 50 600 730, page=4]{Gr_Photoconductivity.pdf}
\includegraphics*[viewport=80 50 600 730, page=5]{Gr_Photoconductivity.pdf}
\includegraphics*[viewport=80 50 600 730, page=6]{Gr_Photoconductivity.pdf}
\includegraphics*[viewport=80 50 600 730, page=7]{Gr_Photoconductivity.pdf}
\includegraphics*[viewport=80 50 600 730, page=8]{Gr_Photoconductivity.pdf}
\includegraphics*[viewport=80 50 600 730, page=9]{Gr_Photoconductivity.pdf}
\includegraphics*[viewport=80 50 600 730, page=10]{Gr_Photoconductivity.pdf}
\includegraphics*[viewport=80 50 600 730, page=11]{Gr_Photoconductivity.pdf}
\includegraphics*[viewport=80 50 600 730, page=12]{Gr_Photoconductivity.pdf}
\includegraphics*[viewport=80 50 600 730, page=13]{Gr_Photoconductivity.pdf}
\includegraphics*[viewport=80 50 600 730, page=14]{Gr_Photoconductivity.pdf}
\includegraphics*[viewport=80 50 600 730, page=15]{Gr_Photoconductivity.pdf}
\includegraphics*[viewport=80 50 600 730, page=16]{Gr_Photoconductivity.pdf}
\includegraphics*[viewport=80 50 600 730, page=17]{Gr_Photoconductivity.pdf}
\includegraphics*[viewport=80 50 600 730, page=18]{Gr_Photoconductivity.pdf}

\title{Supplementary Information: \\
Photoconductivity of biased graphene}% Force line breaks with \\
\author{Marcus Freitag}
\author{Tony Low}
\author{Fengnian Xia}
\author{Phaedon Avouris}
\affiliation{IBM T.J. Watson Research Center, Yorktown Heights, NY 10598, USA
}
\date{\today}

\maketitle

\section{Extracting and correcting for the photo field-effect}

We are interested in the photocurrent that is generated by photons absorbed in the active channel of the graphene photodetector. These photons produce electron-hole pairs in the graphene, which rapidly decay into a cloud of hot electrons and holes, leading to photocurrents due to the photovoltaic, thermoelectric, and bolometric effects. In addition, there exists a photocurrent contribution that is extrinsic to the graphene photodetector, and which we would like to correct for. This contribution is due to light absorbed in the Silicon substrate close to the Si/SiO$_2$ interface, producing a photovoltage at the interface, which is picked up by the gate-sensitive graphene field-effect transistor as a change in source-drain current. It should be possible to avoid this ``photo field-effect'', by using metallic gates, but as we show below, it is also easy to correct for the effect because the intrinsic and extrinsic photocurrent contributions can be spatially decomposed. \\

Due to a workfunction mismatch between Silicon and Silica, the conduction and valence bands in Silicon bend at the interface. For n-type doping of the Silicon substrate as in our case, the bands in Silicon bend upward, which leads to a triangular potential well for holes at the interface \cite{nicollian82}. Photo-generated holes diffuse toward the interface, while electrons are repelled from the interface. This leads to an additional positive voltage on the interface, which acts just like an applied positive gate voltage would in the graphene field-effect transistor, altering the source-drain current. Since the transconductance of a graphene field-effect transistor switches sign at the Dirac point, the photo field-effect also switches sign at the Dirac point ($V_{CNP}\approx 1\,$V in Fig. S1a). This is in contrast to the intrinsic photocurrent, which switches sign twice, as discussed in the main text. \\

The magnitude and spatial extend of the photo field-effect depends on the substrate chemical doping. For intrinsic or lightly doped silicon, the carrier lifetime is long, and the magnitude and spatial extend can be large (centimeters). For heavily doped Silicon, as in our case, the lifetime is shorter, but we still measure a photo field-effect, as can be seen from Fig. S1a, where the photocurrent is plotted as a function of gate voltage and position perpendicular to the graphene channel. The intrinsic photocurrent components decay rapidly once the laser spot moves away from the graphene, but the photo field-effect remains up to a distance of several microns. This behavior allows us to estimate the magnitude of the photo field-effect at the position of the graphene by considering the photocurrent that is generated away from the graphene and fitting it spatially to Lorentzians as exemplified in Fig. S1b. Figure S1e shows the values of the extracted photo field-effect at the center of the graphene as a function of gate voltage. As expected, the curve is proportional to the transconductance $g_m$ extracted from the $I-V_G$ characteristic. The proportionality factor is $2\,$nA/$\mu$S at a laser power of $370\,\mu$W. This means that a photovoltage of 2mV is generated at the Si/SiO$_2$ interface. We can now subtract the photo field-effect component from the total photocurrent and obtain the intrinsic photocurrent in Figs. S1c and S1f. This latter result is used as the basis for our model on the photovoltaic and bolometric components of the intrinsic photocurrent. \\

\section{Spatial distribution of the AC photocurrent and photocurrent saturation at high bias}

The spatial distribution of the photocurrent in biased graphene along the channel direction is shown in Fig. S2 as a function of gate voltage for different drain voltages. At zero drain voltage, the well-known contact effect is present, where regions close to the metallic leads become photoactive because of band-bending there. Both the photovoltaic effect and the Seebeck effect likely play a role in this regime. The contact effect is strongest with the graphene channel electrostatic doping opposite to the metal-induced doping of the graphene beneath the leads, which produces two back-to-back p-n junctions. In our case the metal dopes the graphene n-type and p-n junctions exist for negative gate voltages. These junctions move further into the channel for gate voltages that approach the flat-band voltage at $V_G$=2V. At more positive gate voltages, no p-n junctions exist, and the photocurrent from the contact regions is smaller and is generated right at the contacts.

Once a drain bias in excess of about $V_D$=0.5V is applied, the bias-induced photocurrent, which is the topic of this paper, dominates. The high spatial uniformity of this photocurrent is apparent at $V_D$=-1V, where the middle 4$\mu$m of the 6$\mu$m long graphene shows essentially the same photocurrent and gate-voltage dependence. Contact effects are limited to a 1$\mu$m area next to the metal leads. There is a slight tilt in the gate-voltage characteristic due to drain-voltage induced doping of the channel interior, which affects the right (drain) side of the device more than on the left, and which shifts the photocurrent pattern down by 1V at the drain and half of that (0.5V) in the center of the device. This tilt becomes stronger at $V_D$=-2V and -3V as expected. In fact, one can use these photocurrent measurements to determine the Dirac point inside the biased graphene channel as a function of x-position.

The saturating behavior of the bolometric component of the photocurrent is already becoming apparent below $V_D$=-1V (see main text Fig. 3e). The color-scale bars in Fig. S2 show that at higher drain voltages both the BOL and PV components indeed saturate. Once the electron temperature is elevated due to the bias, additional photogenerated carriers will not be able to increase the electron temperature as much as before, because the photocarrier lifetime will be reduced if the electron distribution is already hot. The high bias thus limits both BOL and PV components of the AC photocurrent.
%OTHER MECHANISMS THAT MIGHT BE IMPORTANT HERE?

\section{Device modeling}

We consider back-gated ($V_{G}$) graphene devices, where the left
contact is grounded i.e. $V_{L}=0$ and $V_{R}$ allowed to vary.
Our model considers the operating regime where the bias current $I_{dc}$
induced by $V_{R}$ is still in the linear regime.
The electrochemical potential $\mu$ in the graphene
channel ($-\tfrac{L}{2}$$<$$x$$<$$\tfrac{L}{2}$) is simply,
\begin{eqnarray}
\mu(x)=\frac{eV_{R}}{L}x-\frac{eV_{R}}{2}
\end{eqnarray}
The electrical potential energy $\Phi(x)$ (or Dirac point energy) is given by,
\begin{eqnarray}
\nonumber
\Phi(x)&=&\frac{\beta_{R}-\beta_{L}}{L}x+\frac{\beta_{R}+\beta_{L}}{2}+\mu(x)\\
\beta_{L/R}&=&-\mbox{sign}(V_G-V_{L/R})\times \hbar v_{f} \sqrt{\tfrac{1}{e}\pi C_{B}\left|V_{G}-V_{L/R}\right|}
\label{chaelec}
\end{eqnarray}
To keep the analytics tractable, we fit the electrical conductivity phenomenologically for electron-hole puddles,
\begin{eqnarray}
\sigma(\epsilon)=\frac{\sigma_{min}}{\Delta^2} \sqrt{\epsilon^{4}+\Delta^{4}}
\label{condeq}
\end{eqnarray}
where $\epsilon$ is defined to be $\epsilon$=$\mu-\Phi$.
$\sigma_{min}$ is the minimum conductivity and
$\Delta$ represents the neutrality region energy width. Both can be simply extracted from the experiments,
through $\left\langle \sigma\right\rangle =\tfrac{1}{L}\int \sigma(\epsilon) dx $.
In our experiments, device physical dimensions are $W\times L=1\times 6\,\mu$m and $t_{ox}=90\,$nm.
The experimentally measured graphene electrical conductivity
is fitted to Eq.\,\ref{condeq}, with best fit values of
$\sigma_{min}=2.3\times 10^{-4}\,$S and $\Delta=75\,$meV.
In our experiment, the extracted effective mobility around the neutrality point
is $\mu=0.27\,$m$^2$/Vs.

\section{Thermoelectric current modeling}

The Seebeck coefficient is computed using the Mott formula\cite{cutler69},
\begin{eqnarray}
{\cal S}_{g}=-\frac{\pi^{2}k_{B}^{2}T}{3e}\frac{1}{\sigma}\frac{d\sigma}{d\epsilon}=
-\frac{\pi^{2}k_{B}^{2}T}{3e}\frac{2\epsilon^{3}}{\epsilon^{4}+\Delta^{4}}
\label{sbco}
\end{eqnarray}
The second equality makes use of Eq.\,\ref{condeq}. Hence, ${\cal S}_g$ for each location in $x$ can be computed.
The photocurrent density (Am$^{-1}$) generated by the thermoelectric effect can be
computed through,
\begin{eqnarray}
J_{TE} = -\frac{\left\langle \sigma\right\rangle}{L} \int_{-\tfrac{L}{2}}^{\tfrac{L}{2}} {\cal S}_{g}(x) \frac{d{{\cal T}_{e/h}}}{dx}  dx
\label{jte}
\end{eqnarray}
As mentioned in the main manuscript, the uniform channel doping
can be rendered asymmetric under an applied drain bias, such that
the effective doping along graphene changes gradually across the two contacts.
This spatial variation in doping is described by
$\Phi(x)-\mu(x)$ (see Eq. \ref{chaelec}),
from which the resulting Seebeck coefficient
can be computed from Eq. \ref{sbco}.\\

Consider photo-excitation in the middle of the graphene channel.
A simplified model for the hot electron/hole temperature profiles due to photo-excitation suffice \cite{song11}:
\begin{eqnarray}
{\cal T}_{e/h}(x)=\frac{\dot{Q}L}{\kappa_{0}}\Lambda(x)+{\cal T}_{0}
\label{tempsim}
\end{eqnarray}
where ${\cal T}_{0}$ is the ambient temperature,
$\dot{Q}$ is the absorbed laser power,
$L$ the device length and $\kappa_{0}$ is the
electronic thermal conductivity,
where $\kappa_{0}$ and $\sigma$ are related through the
Wiedemann-Franz relation.
$\Lambda(x)$ is a triangular function, with maximum
at the middle of the channel i.e. $x=0$ and zero at $x=\pm \tfrac{1}{2}L$.
As discussed in Sec. \ref{sec:pvcur},
${\cal T}_{e/h}(x)$ has a maximum temperature
of $8\,$K in the middle of the channel.
The thermoelectric current
calculated from Eq. \ref{jte} yields
$I_{TE}\approx 4\,$nA at $V_{R}=1\,$V and
when graphene channel is biased near charge neutrality.
This thermoelectric effect is an order
smaller than the corresponding photocurrent
observed in experiment and also has an opposite sign.

\section{Photovoltaic current modeling}
\label{sec:pvcur}

As argued in the main manuscript, the observed photocurrent of $I_{PV}\approx 40\,$nA
in graphene when biased near the charge neutrality point (i.e. $V_{G}=0$) is
due to a photovoltaic contribution.
The photovoltaic current can be modeled by,
\begin{eqnarray}
J_{PV} = \sigma^{*}\xi=\frac{ \sigma^{*}}{L}(\beta_{R}-\beta_{L}-eV_{R})
\end{eqnarray}
where $\sigma^{*}$ is the photoexcited conductivity.
With an applied drain bias of $V_{R}=-1\,$V and source $V_{L}=0\,$V, the calculated channel electric
field (using Eq. \ref{chaelec}) when the device is biased near the charge neutrality point
is $\xi=1.53\times 10^{5}\,$V/m. This yields us $\sigma^{*}=2.6\times 10^{-7}\,$S.
Since $\sigma^{*}$ can be expressed as
$\sigma^{*}\approx q n^{*}\mu^{*}$,
where $n^{*}$ is the photo-induced carrier
density and $\mu^{*}$ the effective mobility of
these excited carriers, where we assumed $\mu^{*}\approx\mu =0.27\,$m$^2$/Vs,
where $\mu$ is inferred from experiments.
We obtain  $n^{*}=6\times 10^{12}\,$m$^{-2}$ at $V_{G}=0$.\\

The photo-induced electron and hole densities at the laser
spot are estimated to be $n^{*}_{e}=n^{*}_{h}\approx g_{f}n^{*}/2$,
where $g_{f}=WL/a_{spot}\approx 16$ is a geometrical scaling factor
with $a_{spot}$ being the focal area.
Hence $n^{*}_{e/h}\approx 5\times 10^{13}\,$m$^{-2}$ at $V_G=0$.
$n^{*}_{e}$ and $n^{*}_{h}$ as function of $V_{G}$ can be modeled with,
\begin{eqnarray}
\nonumber
n^{*}_{e}&=&\int_{0}^{\infty}D(\epsilon)f(\epsilon,{\cal T}_{e}^{1},\mu_{e}^{1})d\epsilon-
\int_{0}^{\infty}D(\epsilon)f(\epsilon,{\cal T}^{0},\mu^{0})d\epsilon\\
n^{*}_{h}&=&\int_{-\infty}^{0}D(\epsilon)[1-f(\epsilon,{\cal T}_{h}^{1},\mu_{h}^{1})]d\epsilon-
\int_{-\infty}^{0}D(\epsilon)[1-f(\epsilon,{\cal T}^{0},\mu^{0})]d\epsilon
\label{nstar}
\end{eqnarray}
and
\begin{eqnarray}
C_{B}V_{G}=e\int_{0}^{\infty}D(\epsilon)f(\epsilon,{\cal T}_{e}^{1},\mu_{e}^{1})d\epsilon
-e\int_{-\infty}^{0}D(\epsilon)[1-f(\epsilon,{\cal T}_{h}^{1},\mu_{h}^{1})]d\epsilon
\label{nstar2}
\end{eqnarray}
where $f$ is the Fermi-Dirac distribution function,
${\cal T}_{e/h}$ and $\mu_{e/h}$ are the respective carrier temperatures and
Fermi levels. The superscript $0$ and $1$ denotes
the absence and presence of light excitation.
$D(\epsilon)=\tfrac{2}{\pi\hbar^{2}v_{f}^{2}}\sqrt{\epsilon^{2}+\epsilon_{0}^{2}}$ is the density-of-states,
where $\epsilon_{0}$ is introduced to account for the electron-hold puddles.
Due to the photo-excitation, the carriers
will be driven away from equilibrium,
characterized by a non-equilibrium Fermi energy $\mu_{e/h}^{1}$
and an elevated carrier
temperatures ${\cal T}_{e/h}^{1}$ compared to the ambient ${\cal T}_{e/h}^{0}$.\\

At steady state, electrons and holes are allowed to thermalize among
themselves, i.e. ${\cal T}_{e}={\cal T}_{h}$ and $\mu_{e}^{1}=\mu_{h}^{1}$,
facilitated by femtosecond time scale carrier-carrier scattering processes \cite{kim11,breusing09}.
Here, ${\cal T}_{e/h}^{1}$ can be described by
${\cal T}_{e/h}^{1}-{\cal T}_{0}=\dot{Q}L/\kappa_{0}$
where $\dot{Q}$ is the absorbed laser power,
$L$ the device length and $\kappa_{0}$ is the
electronic thermal conductivity.
Since $\sigma$ and $\kappa_{0}$ are related through the
Wiedemann-Franz relation,
${\cal T}_{e/h}^{1}-{\cal T}_{0}$ is then proportional to
$1/{\cal T}_{0}\sigma$, where the proportionality
constant is determined to give us
$n^{*}_{e/h}\approx 5\times 10^{13}\,$m$^{-2}$ at $V_G=0$.
This corresponds to ${\cal T}_{e/h}^{1}-{\cal T}_{0}\approx 8\,$K and $12\,$K at
${\cal T}_{0}=300\,$K and $200\,$K respectively.
The photo-excited carriers $n^{*}_{e/h}$ can then be numerically determined
with Eq. \ref{nstar}-\ref{nstar2} by imposing charge conservation $n^{*}_{e}=n^{*}_{h}$.
Having calculated $n^{*}_{e/h}$ as a function of $V_{G}$
then provides us with an estimatation of $J_{PV}(V_{G})$ used in the main manuscript.\\

In our analysis, we have extracted the photo-induced carrier density $n^{*}$
from electrical measurements described above. Alternatively,
one can also estimates the photo-induced carrier density
based on our light excitation condition.
However, uncertainty in various parameters render it less
accurate than the electrical method. Nevertheless, we can
perform estimates of the photo-induced carrier density
based on our light excitation condition.
In our experiments, the laser power is $P=370\,$$\mu$W with focal area
$a_{spot}=\tfrac{\pi}{4}(0.7)^2\,$$\mu$m$^2$.
Light absorption at $\lambda$=690nm (i.e. photon energy $E_{ph}=1.8\,$eV)
in graphene on 90nm SiO$_2$ is $\alpha\approx 2.5\%$.
The photo-induced carrier density can be expressed
as $n_{e/h}^{*}=M\alpha P\tau_{rc}/E_{ph}a_{spot}$,
where $M$ is the carrier multiplication factor and $\tau_{rc}$ is the carrier recombination time. Since
$n_{e/h}^{*}\approx 5\times 10^{13}\,$m$^{-2}$, we estimate that
$M\tau_{rc}\approx 0.6 \,$ps, which seems reasonable \cite{george08}.

\section{Intrinsic electron-phonon lattice heating}

Electron-electron interaction results in an energy equilibration of the electronic system but does not lead to a net energy loss. The dominant energy loss pathways are due to phonons \cite{bistritzer09,tse09,song11,rotkin09}. In particular, electronic cooling in graphene due to intrinsic acoustic/optical phonon scattering processes has been well studied \cite{bistritzer09,tse09}. For example, the electron-lattice energy transfer mediated by acoustic phonons has the following power density (Wm$^{-2}$) given by \cite{bistritzer09},
\begin{eqnarray}
Q_{ac}\approx \frac{D_{ac}^{2}k_{B}}{\hbar\rho_{m}v_{f}^{2}}({\cal T}_{e}-{\cal T}_{L})\frac{1}{\pi}\int dk k^{3}f(\epsilon_{k},{\cal T}_{e},\mu)
\end{eqnarray}
where $D_{ac}\approx 20\,$eV is the acoustic phonon deformation potential and $\rho_m$ is mass density of graphene. For the experimental condition ${\cal T}_{e}-{\cal T}_{L}\approx 10\,$K and undoped graphene, $Q_{ac}$ is only of the order of $10^{2}\,$Wm$^{-2}$. Under some doping and temperature conditions, the optical power density $Q_{op}$ may dominate over its acoustic counterpart \cite{bistritzer09}, however $Q_{op}/Q_{ac}$ is generally $<100$ over the range of experimentally relevant conditions.

\newpage

\begin{figure}[htps]
\centering
\scalebox{0.7}[0.7]{\includegraphics*[viewport=50 50 600 730]{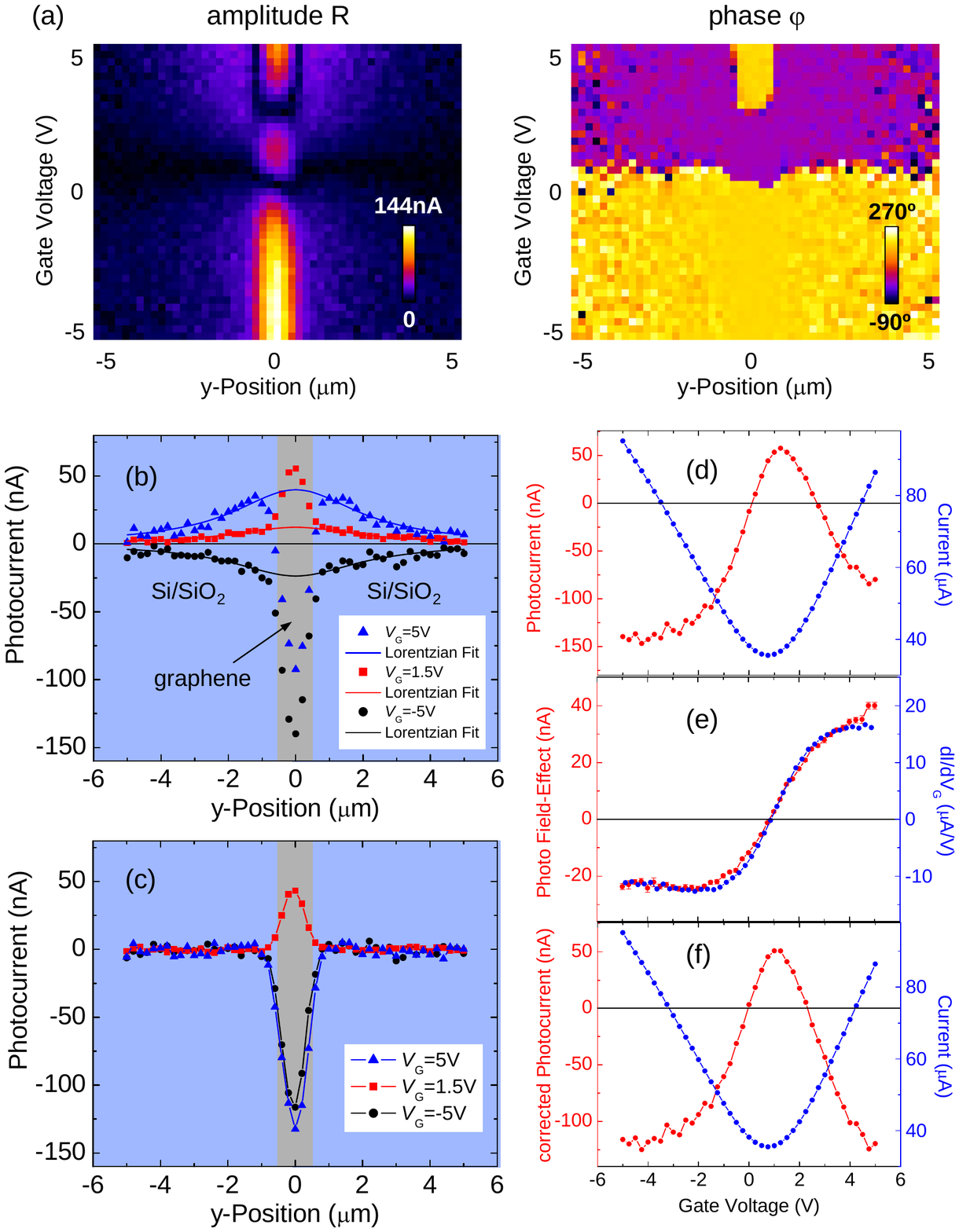}}
\caption{\footnotesize \textbf{(supplemental) Correction for the Photo Field-Effect.} $\bold{(a)}$ Photocurrent amplitude and phase as a function of y-position (perpendicular to the graphene device) and gate voltage. $\bold{(b)}$ Fitting of the photo field-effect component of the photocurrent to Lorentzians for selected gate voltages. The gray-shaded area indicates the position of the 1$\mu$m wide graphene device, which was excluded for fitting purposes. $\bold{(c)}$ Photocurrent as a function of y-position corrected for the photo field-effect for the same gate voltages as in (b). $\bold{(d)}$ Measured AC photocurrent (red) in the center of the graphene FET in Fig. 2 of the main text, and corresponding DC current (blue) as a function of gate voltage. $\bold{(e)}$ Photo field-effect at the center of the graphene channel (red) extracted from fits similar to the ones in (b). The photo field-effect is proportional to the transconductance (blue). $\bold{(f)}$ Photocurrent (red) corrected for the photo field-effect. The DC current (blue) is plotted again as a reference.
}
\label{FigS1}
\end{figure}

\newpage

\begin{figure}[htps]
\centering
\scalebox{0.7}[0.7]{\includegraphics*[viewport=50 350 600 700]{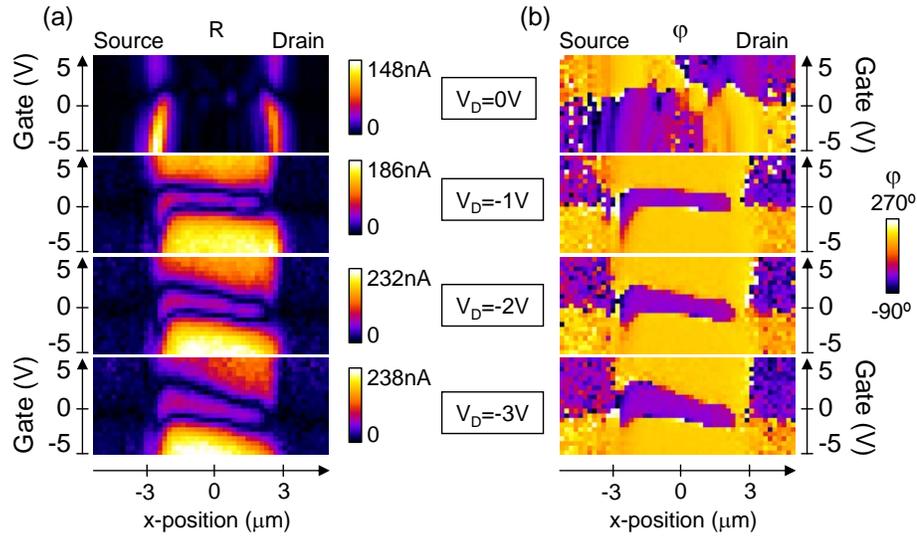}}
\caption{\footnotesize \textbf{(supplemental) Spatial behavior of the AC photocurrent at low and high drain bias.} Amplitude $\bold{(a)}$ and phase $\bold{(b)}$ of the AC photocurrent as a function of gate voltage and x-position along the graphene device for drain voltages from $V_D$=0V to -3V.
}
\label{FigS2}
\end{figure}

\newpage

%---------------------------------------------

\bibliography{ref}% Produces the bibliography via BibTeX.

\end{document}